\documentclass[aps,prd,superscriptaddress,nofootinbib,tighten,preprint]{revtex4}

\newcommand{\be}{\begin{eqnarray}}
\newcommand{\ee}{\end{eqnarray}}

\def\nue{{\nu_e}}
\def\anue{{\bar\nu_e}}
\def\numu{{\nu_{\mu}}}
\def\anumu{{\bar\nu_{\mu}}}

\newcommand{\ms}{\Delta m^2_{21}}
\newcommand{\ma}{\Delta m^2_{31}}

\def\nn{\nonumber}

\def\gtap{\ \raisebox{-.4ex}{\rlap{$\sim$}} \raisebox{.4ex}{$>$}\ }

\newcommand{\epsem}{\epsilon_{e\mu}}
\newcommand{\epset}{\epsilon_{e\tau}}
\newcommand{\epsmm}{\epsilon_{\mu\mu}}
\newcommand{\epsmt}{\epsilon_{\mu\tau}}
\newcommand{\epstt}{\epsilon_{\tau\tau}}

\def\gs{\mathrel{
   \rlap{\raise 0.511ex \hbox{$>$}}{\lower 0.511ex \hbox{$\sim$}}}}
\def\ls{\mathrel{
   \rlap{\raise 0.511ex \hbox{$<$}}{\lower 0.511ex \hbox{$\sim$}}}}
\newcommand{\bea}{\begin{equation} \begin{array}{c}}
\newcommand{\bead}{\begin{equation} \begin{array}{cccc}}
\newcommand{\eea}{ \end{array} \end{equation}}
\usepackage{amsfonts}
\usepackage{amssymb}
\usepackage{amsmath}
\usepackage{graphicx,subfigure}
\usepackage{bbm}
\def\slc#1{\setbox0=\hbox{$#1$}           % set a box for #1
    \dimen0=\wd0                                 % and get its size
    \setbox1=\hbox{/} \dimen1=\wd1               % get size of /
    \ifdim\dimen0>\dimen1                        % #1 is bigger
       \rlap{\hbox to \dimen0{\hfil/\hfil}}      % so center / in box
       #1                                        % and print #1
    \else                                        % / is bigger
       \rlap{\hbox to \dimen1{\hfil$#1$\hfil}}   % so center #1
       /                                         % and print /
    \fi}

\begin{document}

\title{Atmospheric Neutrinos: Status and Prospects}

\author{Sandhya Choubey}
\email{sandhya@hri.res.in}

\affiliation{Harish-Chandra Research Institute, Chhatnag Road, Jhunsi, Allahabad 211 019, India}
\affiliation{Department of Theoretical Physics, School of
Engineering Sciences, KTH Royal Institute of Technology, AlbaNova
University Center, 106 91 Stockholm, Sweden}

\begin{abstract}
We present an overview of the current status of neutrino oscillation studies at 
atmospheric neutrino experiments. While the current data gives some tentalising hints 
regarding the neutrino mass hierarchy, octant of $\theta_{23}$ and $\delta_{CP}$, 
the hints are not statistically significant. We summarise the sensitivity to these sub-dominant 
three-generation effects from the next-generation proposed atmospheric neutrino 
experiments. We next present the prospects of new physics searches such as 
non-standard interactions, sterile neutrinos and CPT violation studies at 
these experiments. 
\end{abstract}

\maketitle

\section{Introduction}

In 1996, data from atmospheric neutrinos at the Super-Kamiokande (SK) experiment 
confirmed neutrino flavor oscillations beyond any doubt \cite{Fukuda:1998mi}.
This established the existence of neutrino masses and mixing, and 
was hailed as the first unambiguous evidence of 
physics beyond the Standard Model (SM) of elementary particles. 
Finally, the year 2015 
witnessed the awarding of Nobel Prize to Professor Takaaki Kajita for leading 
the SK collaboration to this remarkable discovery of 
flavor oscillation of atmospheric neutrinos. 
Professor Kajita shared the Nobel Prize with 
Professor Art McDonald of the Sudbury Neutrino Observatory, 
who was given the award for unambiguously establishing flavor 
oscillations of the solar neutrinos \cite{Ahmad:2002jz}. 

\begin{figure}[h]
\begin{center}
\includegraphics[width=0.5\textwidth]{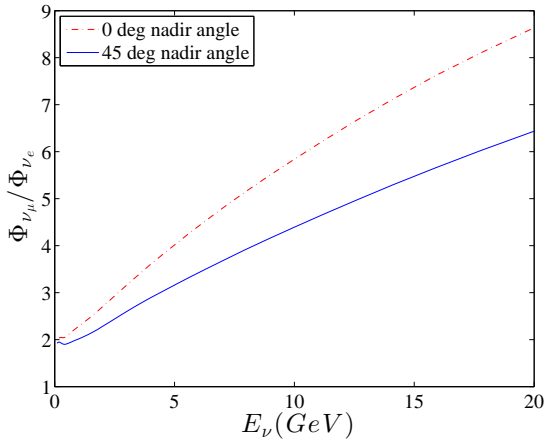}
\caption{Ratio of $\Phi_\numu/\Phi_\nue$ fluxes as a function of 
neutrino energy for two nadir angles of $0^\circ$ and $45^\circ$.
}
\label{fig:flux}
\end{center}
\end{figure}

Atmospheric neutrinos are produced when cosmic ray particles collide with the nuclei 
in the earth's atmosphere, producing pions and kaons which subsequently decay into neutrinos. 
\be
\pi^{\pm} &\to& \mu^{\pm} + \numu(\anumu), \nn \\
\mu^{\pm} &\to& e^{\pm} + \anumu(\numu) + \nue(\anue) \nn
\,.
\ee
We can see that these sets of decay channels would give the flux ratio of muon to 
electron neutrinos of about 2. The exact value of the atmospheric neutrino fluxes 
depend on a lot of issues and are calculated numerically for a given geographical 
location on earth \cite{Honda:2004yz}.
We show in Fig. \ref{fig:flux} this ratio as a function of neutrino energy for 
two neutrino trajectories. The red broken like is for nadir angle of $0^\circ$ (zenith 
angle $180^\circ$) and blue solid line is for nadir angle of $45^\circ$ (zenith 
angle $135^\circ$). We note that this ratio is larger for more vertically traveling 
neutrinos and increases with increasing energy. The reason for the former is that the 
depth of the atmosphere is less for vertical trajectories compared to horizontal 
trajectories, giving vertically traveling particles lesser time to decay. 
The reason for the increase of this ratio with energy is that the higher 
energy particles take longer to decay making the decay process complete 
and leading to fewer electron type neutrinos and antineutrinos. 

Atmospheric neutrinos were originally of interest to the high energy physics community 
mainly because they constituted the most significant background to proton decay 
experiments. Indeed, the first 
observation of atmospheric neutrinos was reported in 1965 at 
the Kolar Gold Field experiment in India \cite{Achar:1965ova}  and almost 
simultaneously by 
an experiment led by Fred Reines in South Africa \cite{Reines:1965qk}, 
both of which were looking for proton decay. 
A discrepancy between the predicted atmospheric neutrino fluxes and that observed 
in detectors was reported first by the Kamiokande II \cite{Hirata:1988uy} 
experiment which was set-up to look for proton decay. 
This anomaly was resolved in terms of flavor oscillations by the 
SK experiment which established the existence of neutrino masses and mixing. 

There are proposals to build bigger and better detectors in the future, some of which 
would be detecting atmospheric neutrinos. 
Amongst the most promising next-generation atmospheric neutrinos 
detectors are the  Hyper-Kamiokande (HK) \cite{Kearns:2013lea}, which will be a 
megaton-class water Cherenkov detector with fiducial volume roughly 20 times that of SK. 
The ICAL detector at the India-based Neutrino Observatory (INO) 
\cite{Ahmed:2015jtv} is proposed to be a 50 kton magnetised iron calorimeter. Being 
magnetised, this detector is expected to have very good charge identification 
efficiency. 
The Precision Icecube Next Generation Underground (PINGU) detector \cite{Aartsen:2014oha} 
is proposed as a low energy extension of the IceCube and is expected to have a 
fiducial volume in the multi-megaton range. This large volume makes this detector 
extremely promising. Along the same lines, with a very large fiducial volume is the 
ORCA proposal which will be the low energy extension of the KM3NeT detector 
in the Mediterranean \cite{Adrian-Martinez:2016fdl}.

In what follows, we will start with a brief discussion of the existing bounds from 
the SK atmospheric neutrino data in section 2. In section 3 we will discuss 
some important aspect of three-generation oscillations. Section 4 and 
5 summarise the expected sensitivity from future experiments for 
neutrino mass hierarchy and octant of $\theta_{23}$, respectively. We discuss 
bounds on non-standard interactions from atmospheric neutrinos in section 
5, sterile neutrinos in section 6 and CPT violation in section 7. We finally 
conclude in section 8.

\section{Neutrino oscillations: Role of atmospheric neutrinos so far}

\begin{figure}[!t]
\begin{center}
\includegraphics[width=0.75\textwidth,angle=90]{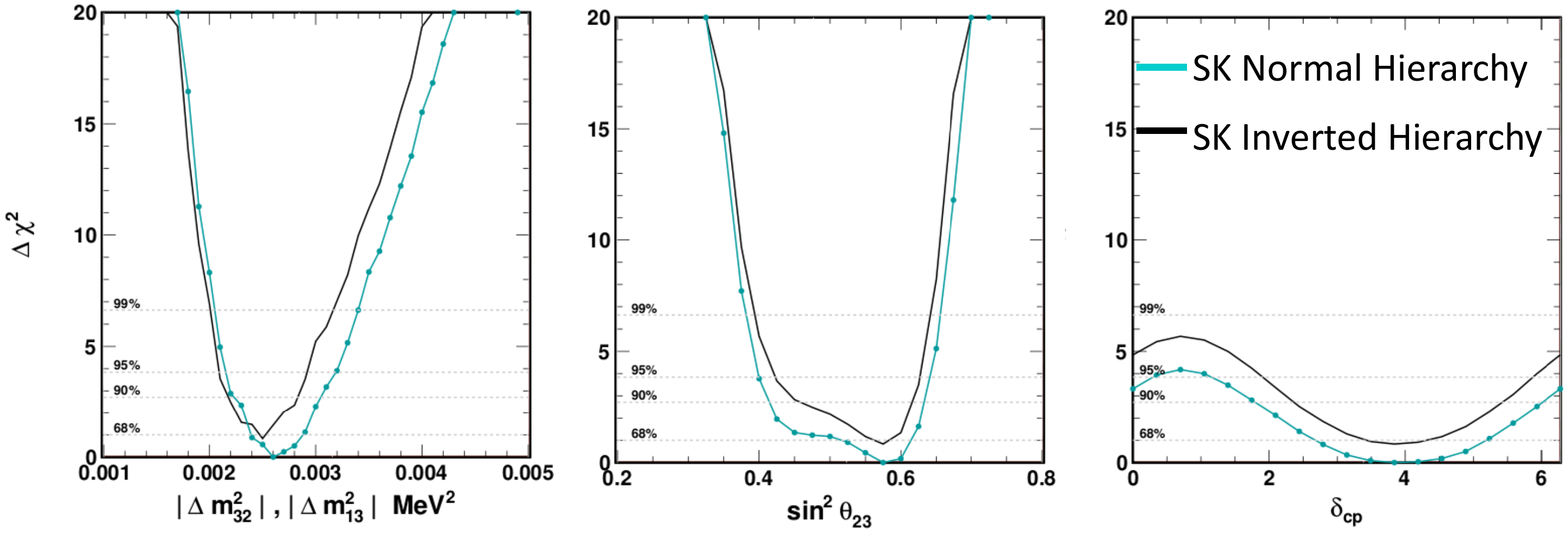}
\vskip -3.2cm
\caption{Current limits from 4581.4 days SK data on atmospheric neutrinos. Reproduced 
from talk by Yoshinaro Hayato on behalf of the SK collaboration, 
at WIN 2015, Germany. 
}
\label{fig:sk}
\end{center}
\end{figure}

The SK experiment until now has collected the most and the best data on atmospheric 
neutrinos. The detector is made of 50 kton of ultra-pure water 
with a fiducial volume of 22.5 kton, and  
started collecting data in April 1996. 
The entire data is divided into 4 sets called SKI, SKII, SKIII and 
SKIV, and the detector continues to operate. 
With 4581.4 days (282.2 kton-yrs) of data analysed, this is statistically 
overwhelming and we summarise the results in Fig. \ref{fig:sk}, 
(taken from talk by Yoshinaro Hayato on behalf of the SK collaboration, 
at WIN 2015, Germany). The left and 
middle-panels of this figure  
show the constraint on the leading atmospheric 
neutrino oscillation parameters $|\ma|$ and $\sin^2\theta_{23}$, respectively. 
The right-panel shows the SK limits on the CP phase $\delta_{CP}$. 
The 
colored lines are for normal hierarchy ($\ma >0$) while the black lines give the 
results for inverted hierarchy ($\ma < 0$). We note that the difference in 
$\chi^2$ between these two cases is not statistically significant. Therefore, this 
implies that the SK data is unable to resolve the sign of $\ma$ even though 
it can constrain its magnitude rather well (cf. left-panel). The weak results 
from the right-panel also indicates that SK is unable to make any statements about the 
CP phase $\delta_{CP}$, though it does give a hint for $\delta_{CP}\simeq 230^\circ$. 
The middle-panel also indicates that the SK data prefers a value of $\theta_{23}$ 
which is non-maximal and also greater than $\pi/4$, however, this hint again 
is not necessarily consistent with other experiments and with global 
analyses of world neutrino data \cite{Gonzalez-Garcia:2015qrr,Capozzi:2016rtj}
and will need confirmation from future experiments. In addition to SK,
we also have recent results on atmospheric neutrinos from the 
MINOS \cite{Adamson:2014vgd}
and IceCube DeepCore \cite{Aartsen:2014yll}. 
When combined with world neutrino data, the leading atmospheric neutrino 
parameters are constrained in the following $3\sigma$ range \cite{Gonzalez-Garcia:2015qrr}
\be
+2.325 \times 10^{-3} < &\Delta m^2_{3l}/{\rm eV}^2& < +2.599\nn \\
-2.59 \times 10^{-3} < &\Delta m^2_{3l}/{\rm eV}^2 &< -2.307 \nn \\
0.385 < &\sin^2\theta_{23}& < 0.644 \nn
\,,
\ee
where $l=1$ for NH and $l=2$ for IH. While the value of $|\ma|$ is 
mainly controlled by the long baseline data from T2K and MINOS, 
$\sin^2\theta_{23}$ is mainly determined by the atmospheric 
neutrino data. 

\section{Three-generation paradigm and the subdominant effects}

\begin{figure}[!t]
\begin{center}
\includegraphics[width=0.32\textwidth]{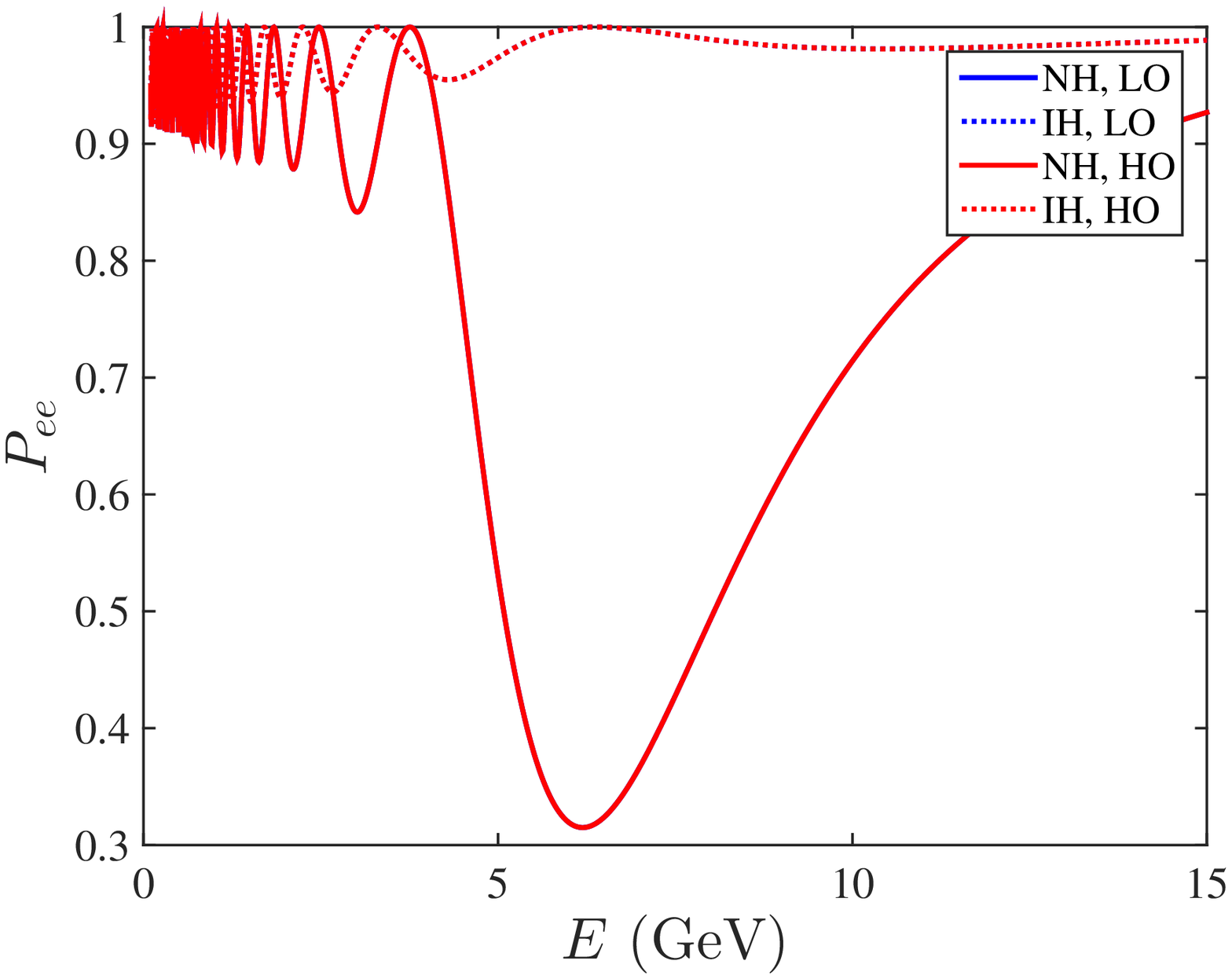}
\includegraphics[width=0.32\textwidth]{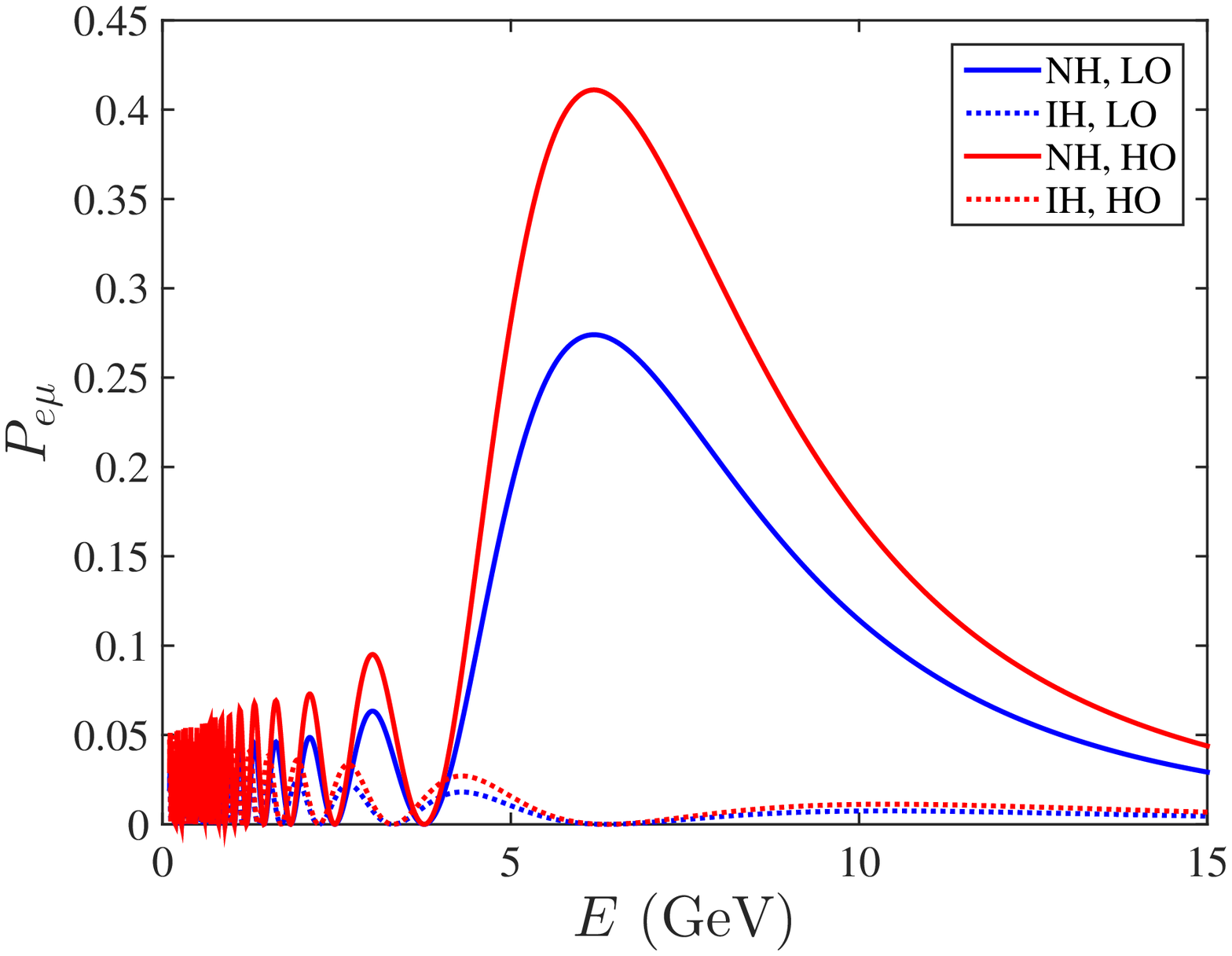}
\includegraphics[width=0.32\textwidth]{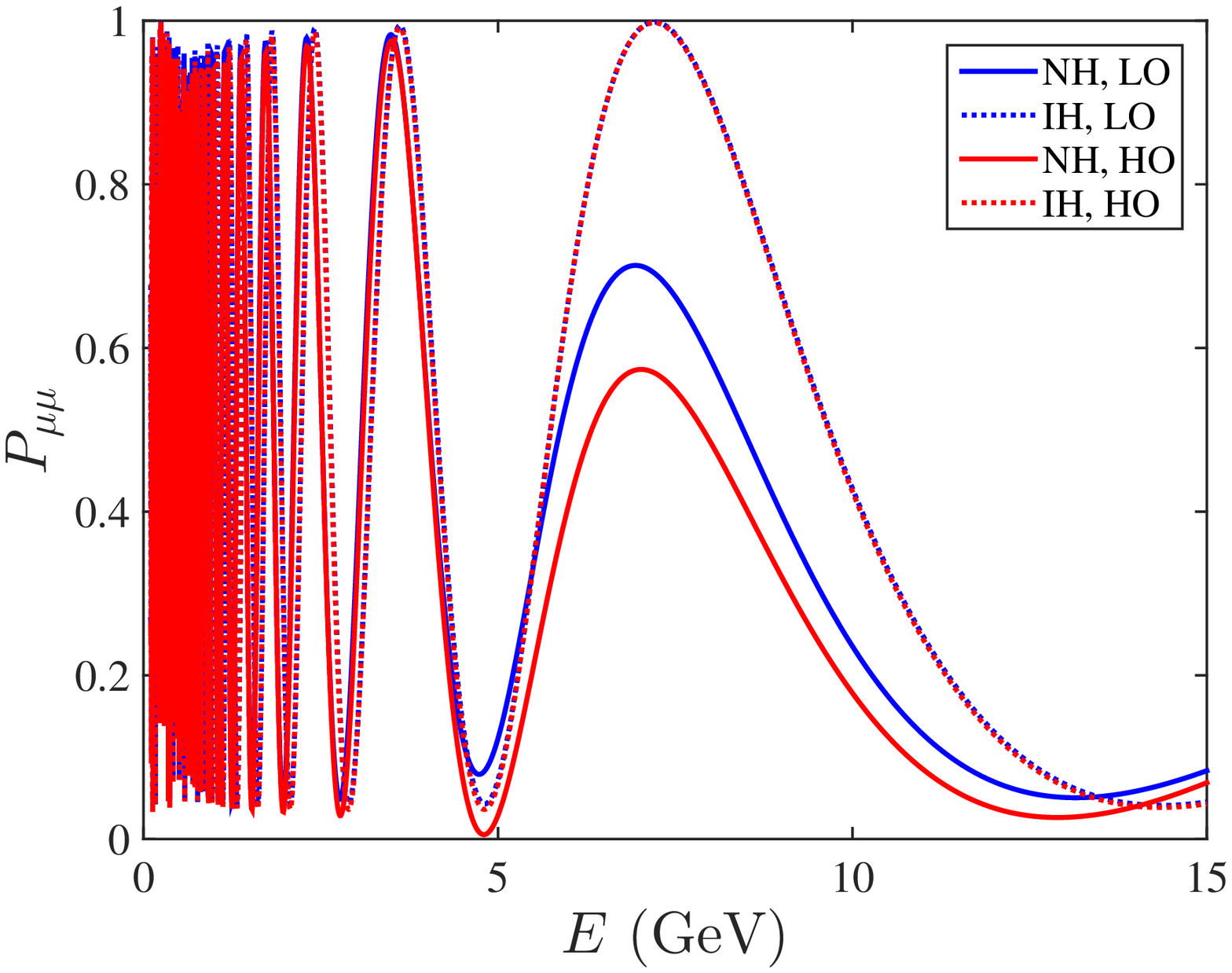}
\caption{Neutrino oscillation probabilities for a baseline of 7,000 km. The solid lines show the 
probabilities for $\ma >0$ and hence 
normal hierarchy (NH) while the broken lines are for $\ma < 0$ and inverted hierarchy (IH). 
The blue lines are for $\sin^2\theta_{23}=0.4$ and hence Lower Octant (LO)
while the red lines are for $\sin^2\theta_{23}=0.6$ and hence Higher Octant (HO). For the 
other oscillation parameters we use the following values: $\ma=2.5\times 10^{-3}$ 
eV$^2$, $\ms=7.6\times 10^{-5}$ eV$^2$, $\sin^2\theta_{13} = 0.023$ and 
$\sin^2\theta_{12}=0.304$. 
}
\label{fig:prob}
\end{center}
\end{figure}

Within the three-generation paradigm, the neutrino mass and mixing is parametrised in 
terms of 3 masses, 3 mixing angles and 3 CP phases, two of which 
are known as Majorana phases. They do not appear 
in the neutrino oscillations 
and show up only in lepton number violating processes such as neutrino-less double 
beta decay. 
Without them, 
the PMNS mixing matrix of the neutrinos is \cite{Maki:1962mu,Pontecorvo:1957qd}
is parametrised as
\be
U_{PMNS} = 
\begin{pmatrix} 1 & 0 & 0 \\
0 & c_{23} & s_{23} \\
0 & -s_{23} & c_{23}\end{pmatrix}
\begin{pmatrix} 
c_{13} & 0 & s_{13}e^{-\delta_{CP}}  \\
0 & 1 & 0 \\
-s_{13}e^{-\delta_{CP} }& 0 & c_{13}\end{pmatrix}
\begin{pmatrix} 
c_{12} & s_{12} & 0 \\
-s_{12} & c_{12} & 0 \\
0 & 0 & 1\end{pmatrix}
\,.
\ee
When neutrinos travel in matter their coherent forward charged current scattering 
off the ambient electrons leads to an extra effective contribution to the neutrino mass 
matrix \cite{Wolfenstein:1977ue,Mikheev:1986wj,Mikheev:1986gs}
\be
H_f = \frac{1}{2E} U_{PMNS} ~diag(0,\ms,\ma)~U_{PMNS}^\dagger 
+ diag(A,0,0)
\,,
\label{eq:hfmatter}
\ee
where $A=\pm \sqrt{2}G_F N_e$ is the effective matter potential
\cite{Wolfenstein:1977ue,Mikheev:1986wj,Mikheev:1986gs}, given in terms of 
the Fermi constant $G_F$ and electron density in matter $N_e$. 
The sign of $A$ is positive for neutrinos and negative for antineutrinos. 
It is seen that 
when GeV energy range atmospheric neutrinos travel inside the earth matter, 
they encounter sizeable changes due to the matter term which depends directly 
on the sign of $\ma$ and the value of $\theta_{13}$.  For $\theta_{13}=0$, the 
matter effect is negligible, but since we now have very strong experimental 
evidence of $\sin^2\theta_{13} \simeq 0.025$ from Daya Bay, RENO, Double 
Chooz, T2K, MINOS and NO$\nu$A (see \cite{Gonzalez-Garcia:2015qrr,Capozzi:2016rtj}
for a global analysis of data from all these experiments, 
and references therein), earth matter effects in atmospheric neutrinos are 
expected to be significant in the neutrino channel for $\ma >0$ and 
in the antineutrino channel for $\ma < 0$. 

While the SK atmospheric data has confirmed the leading $|\ma|$-driven flavor 
oscillations 
beyond doubt, the subdominant effects coming from the three-generation paradigm 
remains to be confirmed, as was discussed in the previous section. 
The most important issues on which the data from future
atmospheric neutrino experiments could throw light are the issue of 
the sign of $\ma$, {\it aka}, the neutrino mass 
hierarchy or the neutrino mass ordering
\cite{Bernabeu:2001xn,GonzalezGarcia:2002mu,Bernabeu:2003yp,PalomaresRuiz:2004tk,Indumathi:2004kd,Gandhi:2004bj,Petcov:2005rv,Choubey:2006jk,Akhmedov:2006hb,Gandhi:2007td,Gandhi:2008zs,Barger:2012fx,Blennow:2012gj,Ghosh:2012px,Ghosh:2013mga,Devi:2014yaa,Blennow:2013vta,Winter:2013ema,Ge:2013ffa}
and the correct octant of $\theta_{23}$
\cite{Choubey:2005zy,Indumathi:2006gr,Samanta:2010xm,Choubey:2013xqa}
meaning whether $\theta_{23} < \pi/4$ or $>\pi/4$. 
In addition, these experiments could also play a role in CP studies 
\cite{Peres:2003wd,Akhmedov:2008qt,Ghosh:2014dba,Ghosh:2015ena}
at long baseline 
experiments which suffer due to their lack of knowledge of the mass hierarchy. 
In most cases, its mainly the presence of the $\theta_{13}$-driven 
earth matter effects which give atmospheric neutrinos the handle to probe these 
issues, although sometimes $\ms$-driven subdominant oscillations are also instrumental 
in the diagnostics. 

We show in Fig. \ref{fig:prob} the neutrino oscillation probabilities as a function of the 
neutrino energies for the earth mantle crossing neutrino trajectory corresponding to 
a baseline of $L=7000$ km. The solid lines show the 
probabilities for $\ma >0$ and hence 
normal hierarchy (NH) while the broken lines are for $\ma < 0$ and inverted hierarchy (IH). 
The blue lines are for $\sin^2\theta_{23}=0.4$ and hence Lower Octant (LO)
while the red lines are for $\sin^2\theta_{23}=0.6$ and hence Higher Octant (HO). 
We see that the probabilities are distinctly different between the NH and IH as well as 
between LO and HO cases. In both cases, its the earth matter effects which bring in 
the major difference between NH vs IH as well as LO vs HO. For the electron neutrino 
survival probability $P_{ee}$ there is no effect of $\theta_{23}$ since it does not 
depend on this parameter, but shows a sharp dependence on the mass hierarchy. 
The muon neutrino survival probability $P_{\mu\mu}$ changes both due to hierarchy as well as 
$\theta_{23}$ octant,  however, the octant effect shows up only for the mass hierarchy 
case which develops earth matter effects. In addition, the difference between the 
probability corresponding to NH and IH also changes sign as we change neutrino 
energy $E$ (as can be seen from Fig. \ref{fig:prob}) as well as $L$ (for which one has to do this 
figure for another baseline). The earth matter effects seen in the 
conversion probability $P_{e\mu}$  from $\nu_e$ to $\nu_\mu$ 
shown in the middle panel depends on the octant and 
does not change sign with L and E. From the discussion above we conclude 
that in order to see earth matter effects in muons through $P_{\mu\mu}$ one needs a detector 
with good energy and angle (and hence $L$) resolution. On the other hand, 
these requirements are not mandatory in detectors which can observe electrons and measure 
$P_{ee}$. However, $P_{\mu\mu}$ also brings in information on octant while 
$P_{ee}$ does not. The conversion channel $P_{e\mu}$ is required for both the 
muon as well as electron channels.

\section{Neutrino mass hierarchy}

Discovery of the neutrino mass hierarchy is the next major goal in the field of neutrino 
physics. Mass hierarchy sensitivity is given in terms of the 
difference between the signal at the detector between the NH and IH cases.  
Experiments such as INO, PINGU and ORCA are being proposed 
with the main goal of determining the neutrino mass hierarchy. 
Mass hierarchy determination studies at atmospheric 
neutrino experiments need to face the challenge coming from two major 
screening effects which reduce the sensitivity of these experiments. We 
discuss them briefly below. 

The first challenge comes from the fact that 
matter effects develop only in the neutrino ($N_\alpha({\rm matter}))$) 
or the antineutrino channel ($N_{\bar\alpha}({\rm matter})$) 
for a given true hierarchy. 
However, both neutrinos as well as antineutrinos 
come mixed together in the atmospheric neutrinos flux at the detector. 
For experiments which can distinguish between the charge of the final 
state lepton such as the magnetised ICAL detector at INO, this challenge 
does not pose any significant problem.
On the other hand, for detectors which can't determine the charge of the particle, 
this complication does result in reducing the sensitivity to measuring earth matter 
effects and hence mass hierarchy. 
Since the total events recorded
at the detector for NH is roughly given by $N_\alpha({\rm matter}) + N_{\bar\alpha}$ and for 
IH is $N_\alpha+ N_{\bar\alpha}({\rm matter}))$, the mass hierarchy sensitivity 
can be estimated in terms of the difference 
\be
\Delta N &\simeq&
(N_\alpha({\rm matter}) + N_{\bar\alpha}) - (N_\alpha+ N_{\bar\alpha}({\rm matter})) \nn\\
&\simeq& (N_\alpha({\rm matter}) - N_{\bar\alpha}({\rm matter})) + (N_{\bar\alpha}
-N_\alpha) \nn
\,,
\ee
where $N_\alpha$ and $N_{\bar\alpha}$ are the number of events in the 
neutrino and antineutrino channel of flavor $\alpha$, and 
we have assumed that there is no effect of matter for the neutrino (antineutrinos) 
for NH (IH), which is not a bad assumption. 
If the atmospheric neutrino flux was same for neutrinos and antineutrinos of all 
flavors, and 
if the interaction cross-section of the neutrinos were same as the interaction 
cross-section of the antineutrinos then the number of events for 
neutrinos and antineutrinos would be identically same. In that case 
$N_\alpha({\rm matter}) = N_{\bar\alpha}({\rm matter})$ and 
$N_{\bar\alpha} = N_\alpha$ and $\Delta N =0$, washing out completely 
the mass hierarchy sensitivity. However, this does not happen because even though at the 
probability level the matter effect in the neutrino channel for NH is the same as the 
matter effect in the antineutrino channel for IH, the fluxes of neutrinos are different 
from the fluxes of the antineutrinos, and more importantly, 
the neutrino-nucleon cross-sections are  
lower for the anti-neutrinos than for the neutrinos. 
Therefore, even detectors with no charge identification capability can be 
sensitive to the neutrino mass hierarchy. 

The next level of difficult due to screening of earth matter effects in atmospheric 
neutrinos comes from the fact the 
atmospheric neutrinos come in both muon as well as electron flavors. 
Therefore, if one is observing the $\numu$ signal in the detector, the 
final fluxes are a combination of the survived $\numu$ flux (disappearance channel) 
and the oscillated $\nue$ flux (appearance channel). 
The net flux at the detector is a combination of the original fluxes folded with the 
relevant oscillation probabilities. For instance, the net $\numu$ flux at the detector 
is given by 
\be
\Phi_\numu = \Phi^0_\numu P_{\mu\mu} + \Phi^0_\nue P_{e\mu} 
\,,
\ee
where $\Phi^0_\numu$ and $ \Phi^0_\nue$ are the fluxes before oscillations. 
A quick look at Fig. \ref{fig:prob} reveals that while matter effect reduces $P_{\mu\mu}$ for the 
neutrinos for NH, it increases $P_{e\mu}$. Therefore, the net impact of matter effects 
in atmospheric neutrinos get partially washed down by adding the so-called appearance 
channel ($\Phi^0_\nue P_{e\mu}$) to the disappearance channel ($\Phi^0_\numu P_{\mu\mu}$). 
Again, 
$\Phi^0_\numu / \Phi^0_\nue \simeq 2$ for lower energies and higher 
for higher energies, the cancellation is not complete and the residual matter effects 
can be used to probe the neutrino mass hierarchy.

The most promising next generation atmospheric neutrino experiments 
that could throw light on the mass hierarchy are PINGU, HK, INO and ORCA. All of these 
detectors are planned to be large and can observe earth matter effects to different 
degrees of efficiency. 
While INO has excellent charge identification capabilities, PINGU, ORCA and HK 
are very big in size. In addition, the water and ice detectors are also sensitive to electrons 
and hence can probe the mass hierarchy in both channels.  
%We give in Table 
%\ref{tab:mh} the mass hierarchy sensitivity in the upcoming experiments, HK, Pingu and 
%INO. 
All of these experiments have made available their mass hierarchy sensitivity.  
The $\chi^2$ quoted have been calculated as follows. Data was generated at certain 
assumed true mass hierarchy and with a certain assumed set of 
values for the other oscillation parameters. This is then fitted with the other hierarchy 
allowing the oscillation parameters to vary in the fit and picking out the smallest 
$\chi^2$ value from the set. We give the number of years of 
data in a given experiment to reach $\chi^2=9$  and we do this for both hierarchies assumed 
as true. 
Since the values of $\theta_{13}$ as well as $\theta_{23}$ directly impacts the 
projected sensitivity and since the sensitivity increases with the increase of both these 
mixing angles, we also mention the values of these mixing angles used in the 
relevant study from which these results have been quoted. For NH true 
and $\sin^22\theta_{13}=0.08$ and $\sin^2\theta_{23}=0.5$, 
HK gives a $\chi^2=9$ with about 4.5 years of data. PINGU promises to give a 
$\chi^2=9$ for the wrong mass hierarchy with about 3 years of data for NH true.  
%for 
%$\sin^22\theta_{13}=0.0x$ 
%and $\sin^2\theta_{23}=0.x$ and NH true. 
On the other hand INO being a much smaller detector would need about 
8.5 years of data for $\chi^2=9$ when $\sin^22\theta_{13}=0.1$ 
and $\sin^2\theta_{23}=0.5$ for NH true. However, if one compares the sensitivity 
expected for true IH, it is seen that while the sensitivity of both HK and PINGU 
go down significantly, the sensitivity for INO remains almost the same as in the 
true NH case. The main reason for this is that INO has excellent charged identification 
capabilities which returns nearly the same sensitivity. 
On the other hand HK and PINGU 
suffer due to the partial washing down of the matter effects by mixing of the 
neutrino and antineutrino signal\footnote{See \cite{Choubey:2013xqa} for a 
detailed discussion on a similar issue concerning the decrease in the octant sensitivity 
of PINGU.}.

\section{Octant of $\theta_{23}$}

\begin{figure}[!t]
\begin{center}
\includegraphics[width=0.5\textwidth]{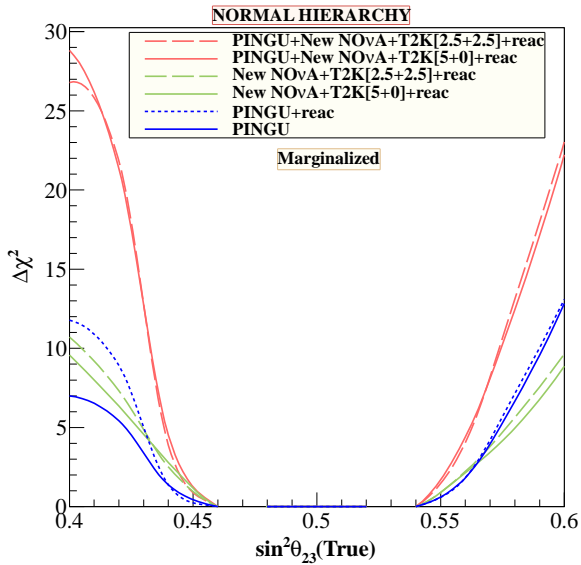}
\caption{$\Delta \chi^2$ as a function of $\sin^2\theta_{23}$(true) for the wrong octant from  
3 years of atmospheric neutrino data from PINGU combined with data from 
3+3 years of running of NO$\nu$A, 5+0 years of running OR 2.5+2.5 years 
of running of T2K and 3 years of running of Daya Bay, RENO and Double Chooz. 
The true hierarchy was assumed to be NH. Reproduced from \cite{Choubey:2013xqa}. 
}
\label{fig:octant}
\end{center}
\end{figure}

We have seen in Fig. \ref{fig:prob} that the earth matter effects in the 
muon neutrino survival probability has a 
$\theta_{23}$ dependence. This gets diluted due to the presence of the appearance channel 
as was also discussed in the earlier section on mass hierarchy. However, the residual 
dependence that remains in the muon data can be used to constrain $\theta_{23}$ 
and find its octant. 
%The prospects of finding the octant in atmospheric neutrino experiments has been 
%explored in \cite{}. 
To illustrate the sensitivity of atmospheric neutrinos to the octant of $\theta_{23}$, 
we show in Fig. \ref{fig:octant} the sensitivity of the PINGU experiment. 
The figure shows the $\Delta \chi^2$ obtained by generating the data for a given 
true value of $\sin^2\theta_{23}$ shown in the x-axis and fitting it with 
values of $\sin^2\theta_{23}$ in the entire wrong octant side and picking out the 
best fit. The blue solid line shows the sensitivity of 3 years of 
PINGU alone, while the blue broken 
line shows the sensitivity when the reactor data is added to PINGU data. The green 
lines show the comparative sensitivity expected from the T2K and NO$\nu$A 
experiments which are seen to have a sensitivity comparable to that of PINGU. 
For NO$\nu$A the simulation uses the fluxes and run times given in their 
DPR while for T2K 2 cases are displayed, one with 5 years of neutrino running 
alone and another with 2.5 years of neutrino and 2.5 years of antineutrino 
running. 
Finally, the red lines show the combined $\chi^2$ obtained when we 
add the T2K, NO$\nu$A and reactor data to the data from 3 years of PINGU. 
The $\chi^2$ for all cases is marginalised over $\ma$ and $\sin^2\theta_{13}$ and 
for T2K and NO$\nu$A we also marginalise over $\delta_{CP}$. 
The figure was generated for NH as true and a similar figure for IH can be found in 
\cite{Choubey:2013xqa}. 
The sensitivity for IH is a little lower and is explained in \cite{Choubey:2013xqa}.

\section{Non-Standard Interactions}

It is now well established that the standard model of particle physics (SM) needs to be extended. 
Most of such extensions of the SM involve addition of new particles and/or extension of 
the gauge sector. The low energy limit of such theories can be expected to have effective 
couplings which are different from that given in the SM. These effective couplings could give 
rise to addition charged current interactions as well as neutral current 
interactions and are in general referred to as Non-Standard Interactions (NSI)
\cite{Wolfenstein:1977ue,Valle:1987gv,Guzzo:1991hi,Roulet:1991sm,Grossman:1995wx}. 
The additional charged interactions due to new physics would 
show up in the production and detection of the neutrinos, 
while the neutral current interaction could significantly impact the neutrino 
propagation inside matter. 
The couplings that drive the additional 
charged current interactions of the neutrinos will also lead to corresponding beyond SM 
interactions in the charged lepton sector due to the SU(2)$_L$ symmetry of the SM. These 
couplings are therefore severely constrained from existing data \cite{Biggio:2009nt}. 
On the other hand, constraints on the neutral current couplings 
are less and have been calculated to be \cite{Biggio:2009nt}
\be
|\epsilon_{\alpha\beta}| < 
\begin{pmatrix}
4.2 & 0.33 & 3.0 \\ 0.33 & 0.068 & 0.33 \\ 3.0 & 0.33 & 21  
\end{pmatrix}
\,,
\label{eq:bound}
\ee
where the parameters have been arranged in the form of a matrix with the rows and columns 
corresponding to $\{ e,\mu,\tau\}$ and the bounds are obtained at the 90\% C.L.. 
We note that the bounds on some of the parameters are 
rather weak. In particular the bound on $\epstt$ looks very loose. These bounds are 
called indirect bounds in the literature since they are derived from non-neutrino-oscillation 
experimental data. 
These parameters also affects the propagation of neutrinos 
in matter through effective operators. 
% of the form 
%\begin{equation}
%{\cal L}_{\rm NSI} = -2\sqrt{2}G_F\epsilon_{\alpha\beta}^{fC}(\overline{\nu_\alpha} \gamma^\mu P_L\nu_\beta)(\overline{f} \gamma_\mu P_C f) \,,
%\end{equation}
%where....
In particular, within the framework of an effective two-generation 
hybrid model, the SK collaboration has put 
a 90\% C.L. limit on the parameter $\epsmt<0.033$ and $|\epstt-\epsmm| < 0.147$ 
\cite{Mitsuka:2011ty,Ohlsson:2012kf}. A comparison with Eq. (\ref{eq:bound}) shows that 
neutrino experiments have much better handle on these parameters and an attempt to 
look at the potential on future experiments to constrain these parameters is pertinent. 
%The potential of 
%some of the proposed long baseline experiments in constraining NSI 
%parameters can be found in \cite{}.
Here we review the prospects of atmospheric neutrino experiments in constraining 
these NSI. In 
addition, since the source-detector NSI are very severely constrained while matter 
NSI are still loosely bounded, in what follows we will discuss the 
current and expected bounds coming from the matter NSI only. 
A lot of work has been done in this field (for illustration, see  
\cite{Fornengo:2001pm,GonzalezGarcia:2004wg,Friedland:2004ah,Friedland:2005vy,GonzalezGarcia:2011my,Escrihuela:2011cf,Ohlsson:2013epa,Esmaili:2013fva,Choubey:2014iia,Mocioiu:2014gua,Chatterjee:2014gxa,Fukasawa:2015jaa,Choubey:2015xha} and references therein)

The neutrino oscillation probabilities change in the presence of NSI. 
The
neutrino oscillation probabilities in the framework on three-generations of neutrinos and 
in presence of NSI were calculated in \cite{Kopp:2007ne}. Keeping first order terms in the 
(small) NSI parameters 
and zeroth terms in $\ms/\ma$ and $\sin^2\theta_{13}$ one obtains 
the difference in the probabilities \cite{Kopp:2007ne,Choubey:2014iia}
\be
\Delta P_{\mu\mu} &=& P_{\mu\mu}^{\rm NSI} - P_{\mu\mu}^{\rm SM} \\ \nn
&\simeq& - |\varepsilon_{\mu\tau}| c_{\phi_{\mu\tau}} \bar A \left[ \sin^3 (2\theta_{23}) \Delta \sin (\Delta) + 4 \sin (2\theta_{23}) \cos^2 (2\theta_{23}) \sin^2 (\Delta/2) \right] \nonumber\\
&+& (|\varepsilon_{\mu\mu}| - |\varepsilon_{\tau\tau}|) \bar A \sin^2 (2\theta_{23}) \cos(2\theta_{13}) \left[ \Delta \sin(\Delta)/2 - 2 \sin^2 (\Delta/2)\right],
\label{eq:deltaPmm_approx}
\ee
where $\Delta \equiv \Delta m^2_{31} L/(2E) $, $\bar A \equiv \sqrt{2} G_F N_eE/\ma
=A/ \ma$, where $A$ is the matter potential defined before. 
A similar expression can be obtained for the other oscillation 
parameters, $\Delta P_{ee}$ and $\Delta P_{e\mu}$, however, we do not given them 
here for the sake of brevity. 

As can be see from Eq. (\ref{eq:deltaPmm_approx}), the muon neutrino 
survival probability predominantly depends on the NSI parameters $\epsmt$ and 
$|\epstt - \epsmm|$. Likewise the conversion channel $P_{e\mu}$ 
primarily depends on the NSI parameters $\epsem$ and $\epset$.  
Constraints on matter NSI from existing atmospheric data have been 
studied in \cite{Fornengo:2001pm,GonzalezGarcia:2004wg,Friedland:2004ah,Friedland:2005vy,GonzalezGarcia:2011my,Escrihuela:2011cf,Esmaili:2013fva,Fukasawa:2015jaa}. 
The NSI parameters are expected to be much better constrained in future atmospheric 
neutrino experiments using bigger and better detectors. 
With three years of data,  we expect the following constraints on the leading NSI 
parameters from PINGU 90\% C.L.($3\sigma$) for NH \cite{Choubey:2014iia}
%\vspace{-5mm}
\begin{align}
-0.0043~(-0.0048) &< \varepsilon_{\mu\tau} < 0.0047~(0.0046) \,, \nonumber\\
-0.03~(-0.016) &< \varepsilon_{\tau\tau} < 0.017~(0.032) ,\nonumber
\end{align}
while INO with 10 years of data could give the following constraints 
at 90\% C.L.($3\sigma$) for NH \cite{Choubey:2015xha} 
\be
-0.12~(-0.28)& <  \epsem <  &0.104~(0.23) \,, \nn \\
-0.13~(-0.3) &<  \epset <  &0.102~(0.21) \,, \nn \\
-0.015~(-0.027) &<  \epsmt <  &0.015~(0.027) \,, \nn \\
-0.07~(-0.104)& <  \epstt <  &0.07~(0.104) \,. \nn 
\ee
The limits for IH come out to be similar. 

\begin{figure}[h]
\begin{center}
\includegraphics[width=0.495\textwidth]{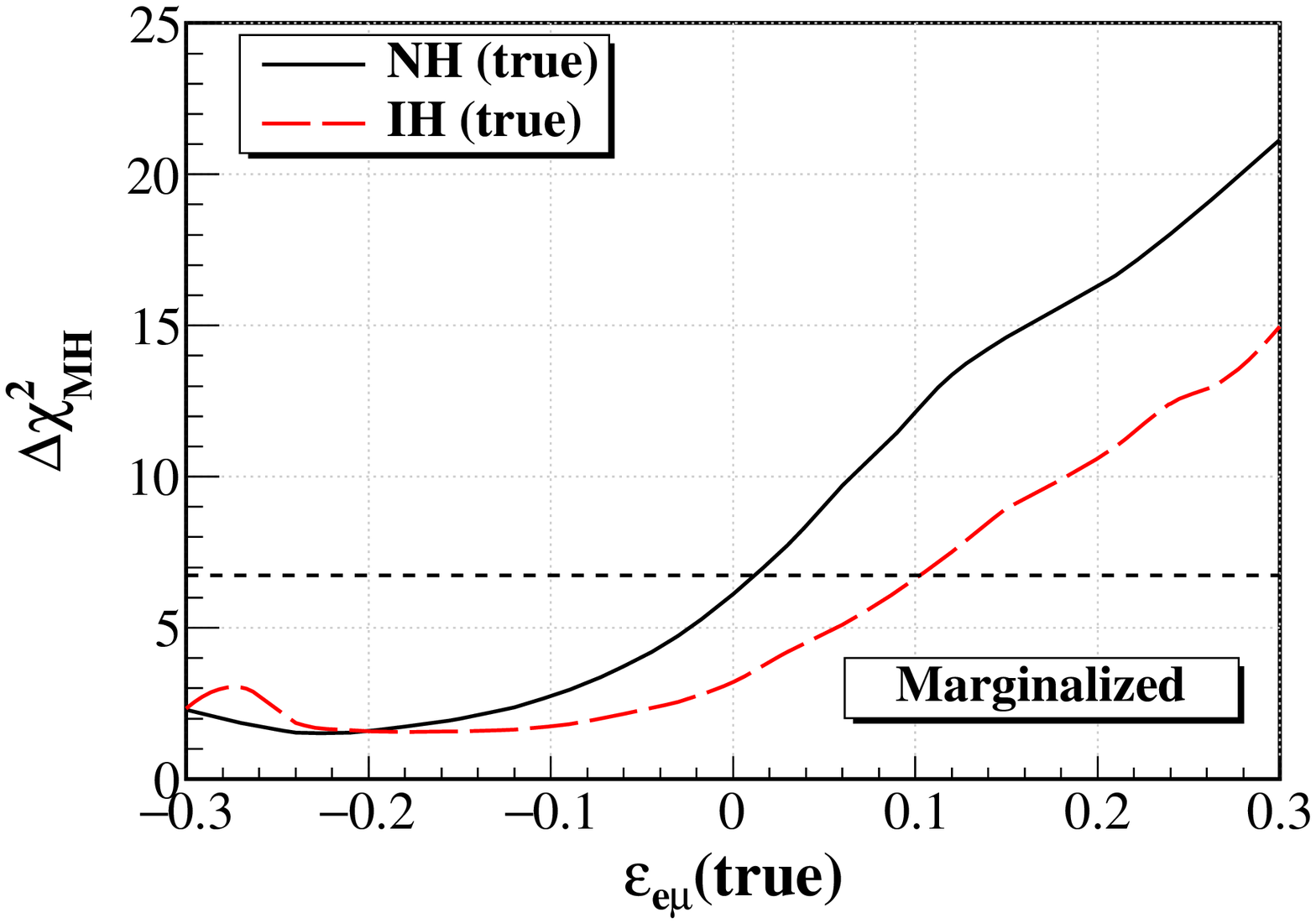}
\includegraphics[width=0.495\textwidth]{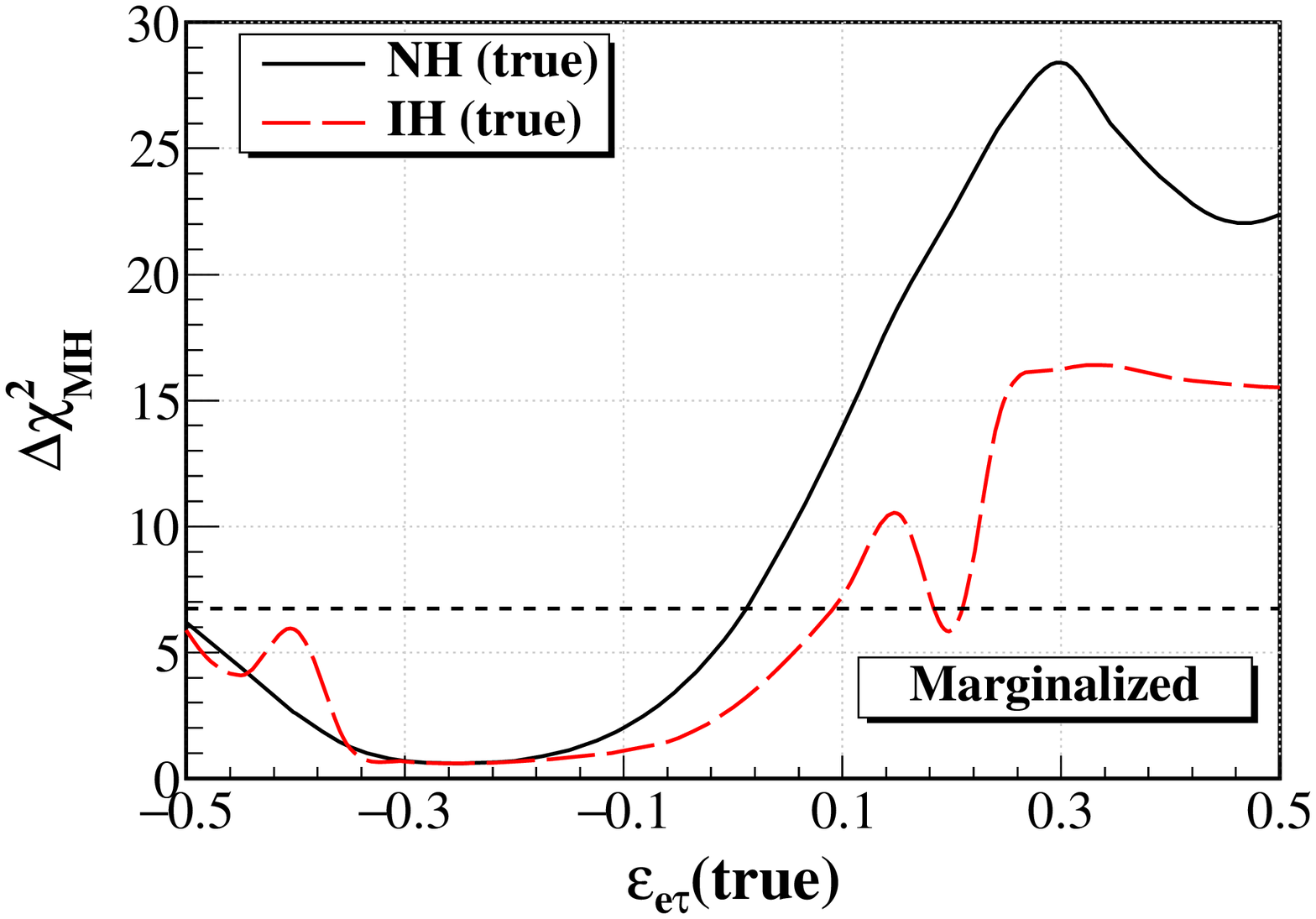}
\includegraphics[width=0.495\textwidth]{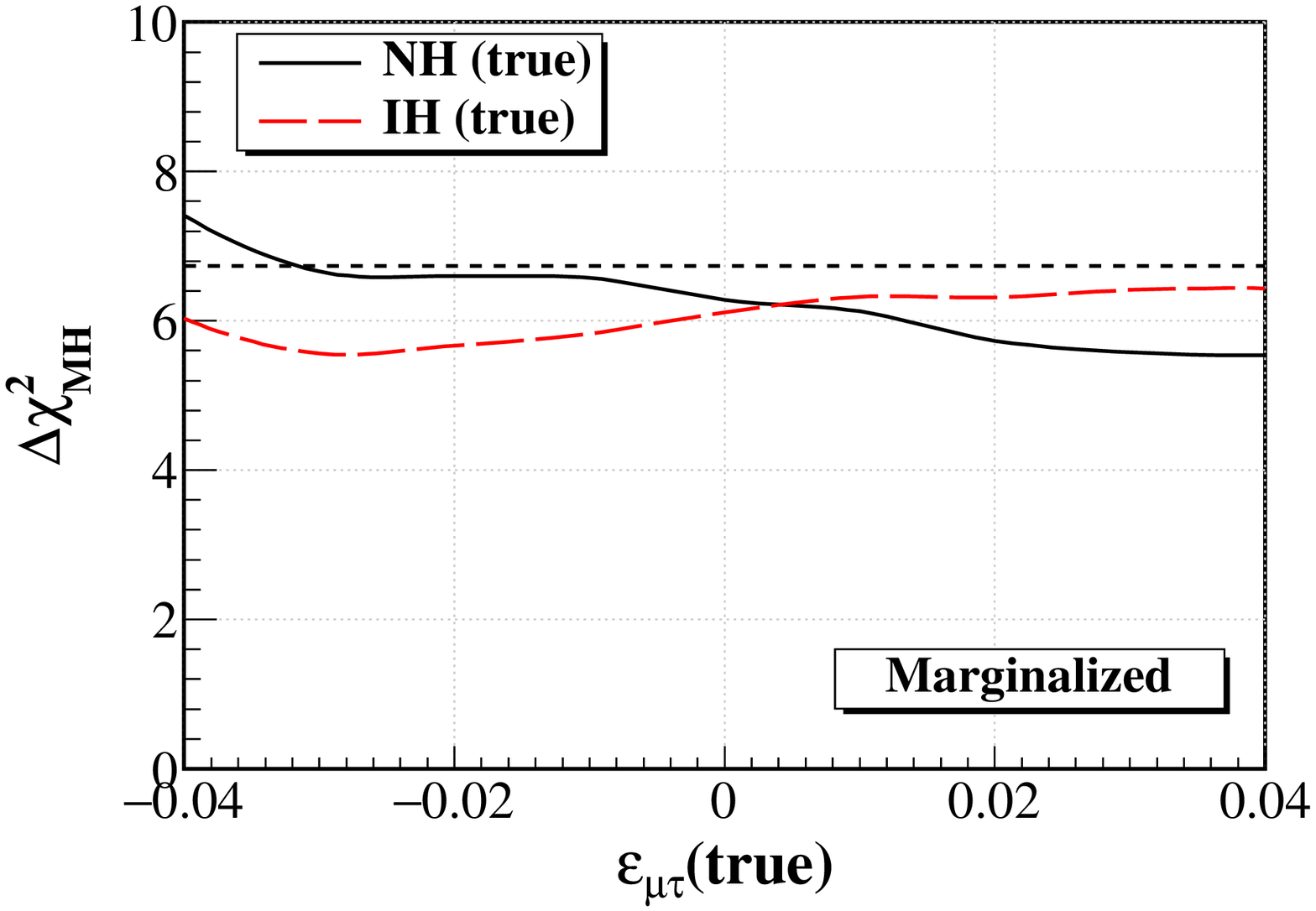}
\includegraphics[width=0.495\textwidth]{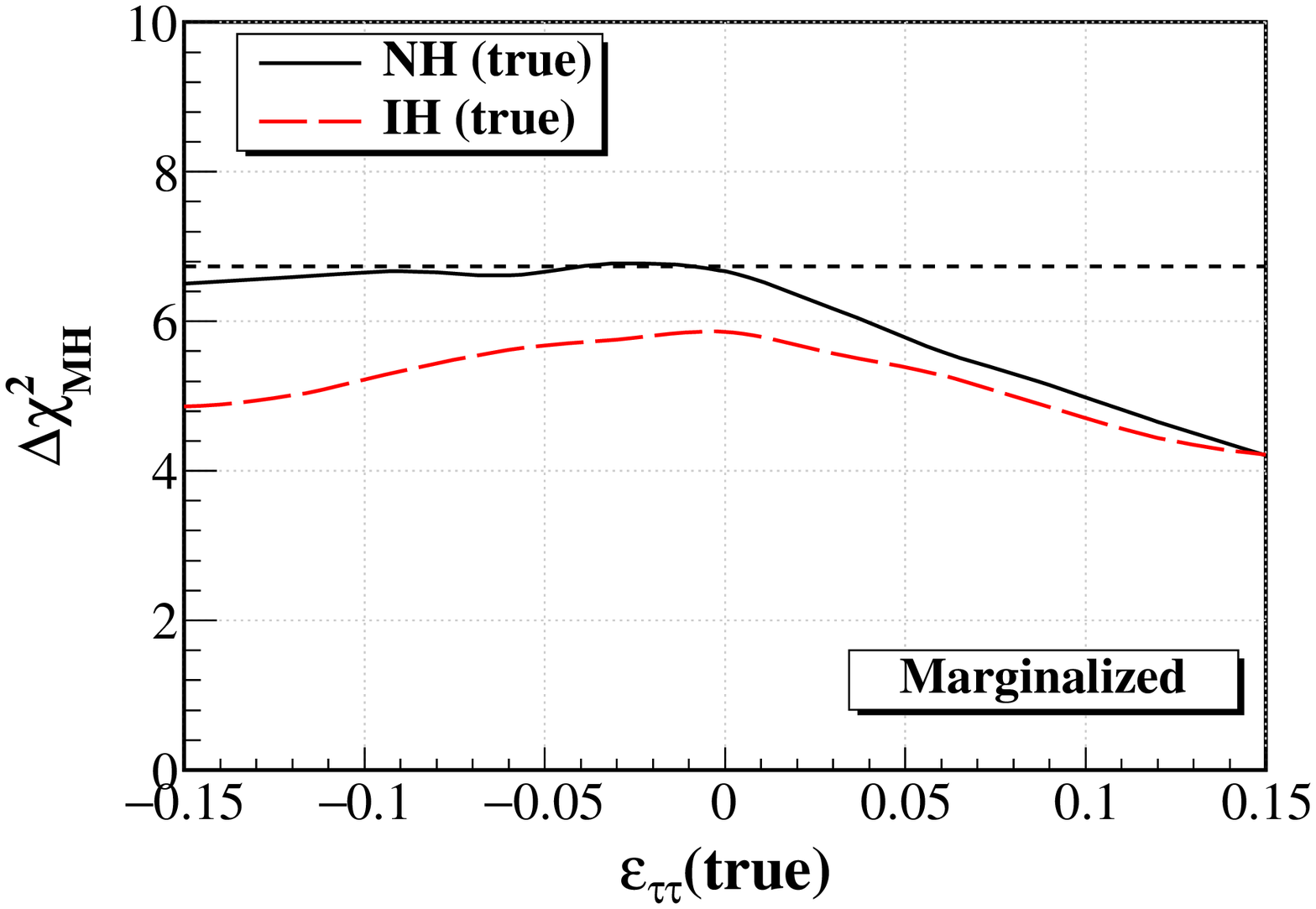}
\caption{The $\Delta \chi^2_{\rm MH}$, giving the 
expected mass hierarchy sensitivity from 10 years of running of INO, 
as a function of the true value of NSI parameters. 
We keep only one $\epsilon_{\alpha\beta}$(true)
to be non-zero at a time, while others are set to zero. 
The $\Delta \chi^2$ is 
obtained after marginalisation over the oscillation parameters.  
Reproduced from \cite{Choubey:2015xha}.
}
\label{fig:nsimh}
\end{center}
\end{figure}

As discussed above, one of the major physics goals of the atmospheric neutrino experiments 
is to determine the neutrino mass hierarchy. For this what is relevant is the difference in 
the neutrino oscillation probabilities between NH and IH which is mainly driven by earth 
matter effects. The presence of matter NSI modifies the effective interaction of the 
neutrinos with matter, changing the earth matter effect and hence the oscillation probabilities. 
This change is different for the two mass hierarchies. The difference in the neutrino oscillation 
probabilities between NH and IH in presence of NSI parameters has been discussed in 
\cite{Choubey:2014iia,Mocioiu:2014gua,Chatterjee:2014gxa} and studied in details in 
\cite{Choubey:2015xha} in the context of INO. We show in Fig. \ref{fig:nsimh} the 
effect of NSI parameters on the mass hierarchy sensitivity of INO. This 
figure given the $\chi^2$ when the data is generated for a given neutrino mass hierarchy and 
a non-zero value of the NSI parameter and the fitted with 
the wrong hierarchy. The corresponding $\chi^2$ is plotted as a function of the 
NSI parameter value used in the data. Only one NSI parameter is taken in the data at a time 
for simplicity. The figure shows that the mass hierarchy sensitivity does change in 
presence of NSI parameters $\epsilon_{e\mu}$ and $\epsilon_{e\tau}$, while 
$\epsilon_{\mu\tau}$ and $\epsilon_{\tau\tau}$ do not affect it much.

\section{Probing sterile neutrinos}

\begin{figure}[!t]
\begin{center}
\includegraphics[width=0.45\textwidth]{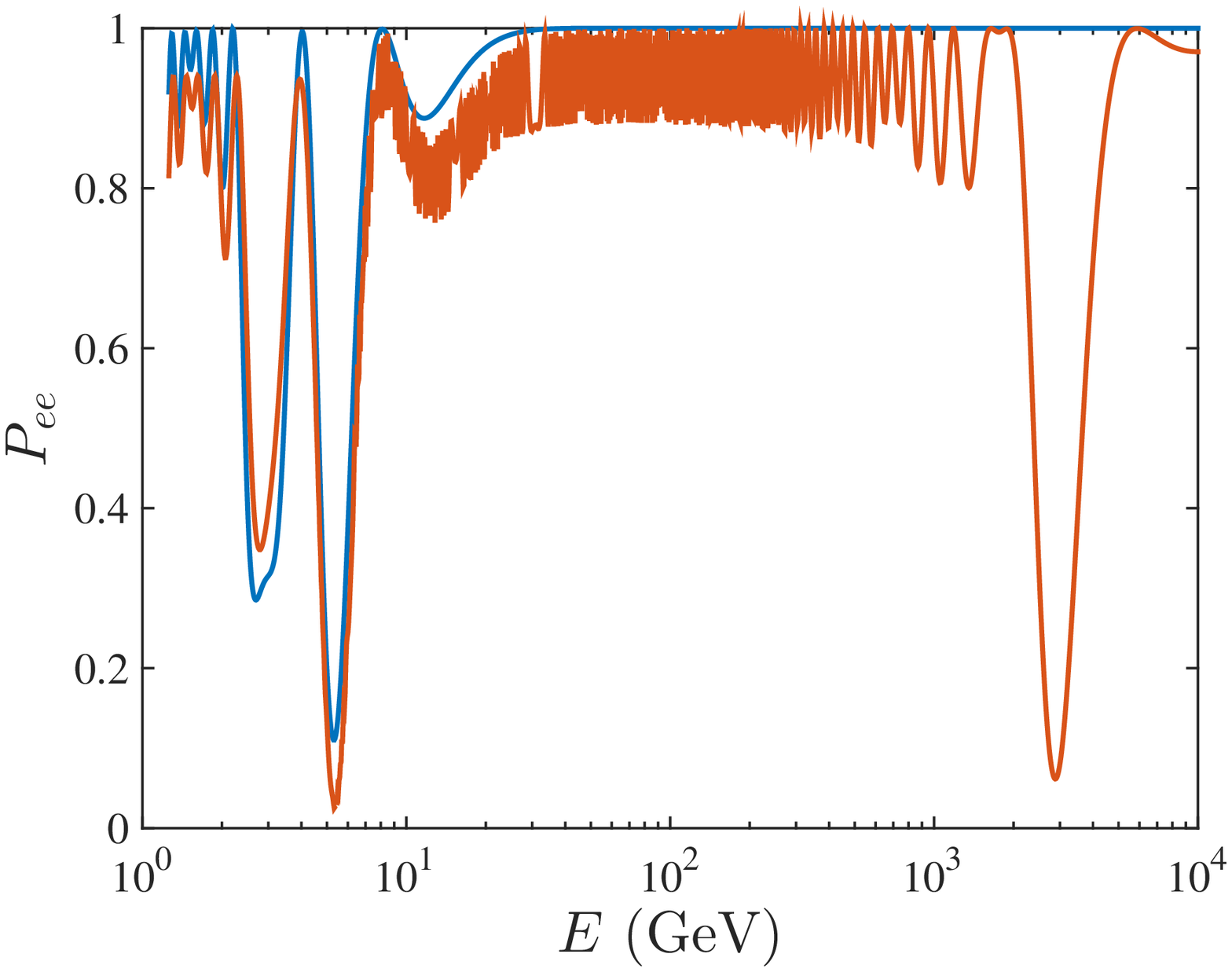}
\includegraphics[width=0.45\textwidth]{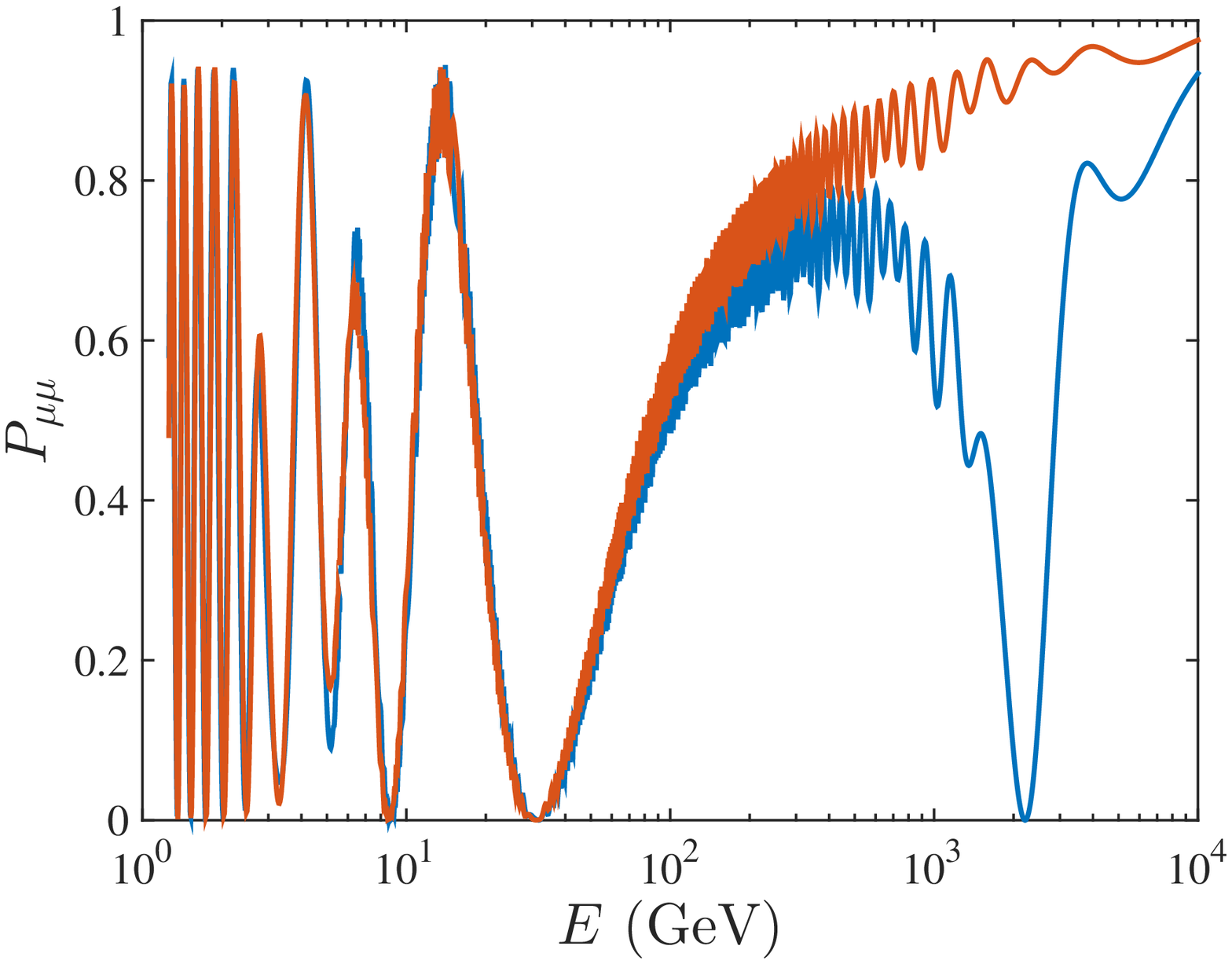}
\caption{Neutrino oscillation probabilities in the 3+1 scenario with 1 extra 
sterile species for the earth centre crossing trajectory. 
The red lines show the survival probabilities 
with mass squared difference $\Delta m^2_{41}=-1.0$ eV$^2$ while the red lines are for 
$\Delta m^2_{41}=1.0$ eV$^2$. All the sterile mixing angles $\theta_{14}$, 
$\theta_{24}$ and $\theta_{34}$ are taken at a benchmark value of $10^\circ$. 
For the other oscillation parameters we use the same 
values as in Fig. \ref{fig:prob}.
}
\label{fig:sterile}
\end{center}
\end{figure}

The decay width of the Z boson measured at 
LEP restricts the number of light neutrino species which couple to the Z 
boson to be very close to 3. 
Hence any light neutrino species beyond the already known three neutrinos should not 
have any gauge interactions and are hence known as sterile neutrinos. 
These additional sterile neutrinos have been postulated as 
a possible explanation of the excess observed at the LSND \cite{Athanassopoulos:1996jb}
and MiniBooNE \cite{Aguilar-Arevalo:2013pmq} experiments. 
In addition to the LSND (and MiniBooNE) hints, we also have the reactor anomaly wherein the 
measured reactor anti-neutrino fluxes are found to be lower than that predicted by the 
theory \cite{Huber:2011wv,Mention:2011rk}, 
and again this discrepancy can be explained by flavor oscillations of sterile neutrinos. 
A variety of accelerator and reactor based experiments have been proposed to 
verify these two anomalies and to confirm or disprove the existence of sterile neutrinos. 

Presence of sterile neutrino species is also expected to alter the flavor 
oscillations of atmospheric neutrinos. This change is brought about due to two reasons. 
Firstly the active-sterile mass squared difference is postulated to be in the 1 eV$^2$ 
regime. This would give rise to very fast oscillations of GeV-range neutrinos, and 
would lead to flavor oscillations of downward neutrinos, which in the three-flavor set-up 
remain unaffected. Secondly, while all active neutrino species undergo neutral current 
scattering over the ambient earth matter, the sterile neutrinos do not interact. This leads to 
neutral current driven matter effects which changes the neutrino mass and mixing 
inside the earth matter and hence oscillation probabilities. 

In Fig. \ref{fig:sterile} 
we show the neutrino oscillation probabilities as a function of the neutrino energies for 
the earth centre crossing neutrino trajectory ($L=2R_e$, where $R_e$ is the 
radius of the earth) in the so-called 3+1 framework, where 
we add 1 extra sterile neutrinos to the 3 active ones. The mixing matrix is 
likewise extended to become $4\times 4$ and for this figure we have used the 
parametrisation 
\be
U_{4gen} = R(\theta_{34})R(\theta_{24})R(\theta_{23})R(\theta_{13})R(\theta_{12})
\,,
\ee
where $R_{ij}$ are the rotation matrices and $\theta_{ij}$ the corresponding 
mixing angle. For neutrino traveling in matter there comes in a further contribution to 
the flavor mixing from the neutral current scattering since while the active flavors all undergo 
(equal) amount of coherent forward scattering, the sterile species remain unaffected. 
Thus the neutrino mass matrix in matter for sterile neutrinos is extended to
\be
H_f = \frac{1}{2E} U_{4gen}~ diag(0,\ms,\ma,\Delta m^2_{41})~U_{4gen}^\dagger 
+ diag(A,0,0,0) + diag(0,0,0,A_{NC} )
\,,
\label{eq:hfmattersterile}
\ee
where $A_{NC} = \pm \sqrt{2}G_FN_n/2$ and we have subtracted out the common neutral 
current components from the active sector leaving behind the term in the sterile part. 
In the way we have written, $A_{NC}$ is positive for neutrinos and negative for antineutrinos. 
Just like in the case of $A$, we encounter matter enhanced resonance
due to $A_{NC}$ as well, as can be seen in the Fig. \ref{fig:sterile}. 
However, unlike the case of active neutrinos, the resonance for sterile 
neutrinos occurs in the $\numu$ sector for $\Delta m^2_{41} <0$ and in the 
$\anumu$ sector for $\Delta m^2_{41} <0$. This is because 
in the effective two-generation approximation, the resonance condition for the 
muon neutrinos is given by $A_{NC} = -\Delta m_{41}^2 \cos\theta_{24}$. For the 
$\nue$ the resonance condition is $A-A_{NC} = \Delta m_{41}^2 \cos\theta_{14}$ and 
hence continues to occur for $\nue$ when $\Delta m^2_{41} > 0$ and for 
$\anue$ when $\Delta m^2_{41} <0$, since $A_{NC} \simeq A/2$. 

From the figure, we see that the earth matter effects changes 
the muon neutrino survival probability somewhat in the energy range 1-100 GeV. This  
energy range can be probed in detectors like SK, INO, PINGU and ORCA. 
The analysis of the current SK data confirms the 
three-flavor paradigm and constrains the sterile neutrino mass and mixing.  
An analysis of the 4438 days of SK atmospheric neutrino data restricts the 
active-sterile neutrino mixing to $|U_{\mu 4}|^2 < 0.041$ and $|U_{\tau 4}|^2 < 0.18$
for $\Delta m_{41}^2 > 0.1$ eV$^2$ at the 90\% C.L. in the 3+1 scenario 
\cite{Abe:2014gda}. 

However, as Fig. \ref{fig:sterile} shows, 
very dramatic effects appear in the oscillation probabilities for the energy 
range of 100 GeV to 10 TeV, where we witness resonant oscillations due to the sterile 
neutrino $\Delta m^2$-driven matter resonance. This dramatic effect on 
atmospheric neutrino fluxes in the TeV energy range has been studied in detail 
in \cite{Nunokawa:2003ep,Choubey:2007ji,Razzaque:2011ab}.
It was pointed out that the TeV atmospheric neutrino events recorded in IceCube 
can be used to constrain the sterile neutrino mass and mixing plane. There have been 
attempts to analyse the IceCube data to constrain the sterile neutrino mixing 
by looking for the sharp change in the expected track to shower ratio 
at IceCube 
\cite{Barger:2011rc,Razzaque:2012tp,Esmaili:2012nz,Esmaili:2013vza,Lindner:2015iaa}.

\section{CPT violation studies}

Lorentz and CPT invariance are one the basic postulates of modern day quantum field theory. 
However, motivated partly by string theories and other attempted 
quantum theories of gravity, there have been some interest in looking for 
breakdown of these symmetries at the Plank scale. 
While a discussion of these theories is outside the scope of this article, we will here 
look at some of the phenomenological implications for atmospheric neutrinos if CPT 
and/or Lorentz invariance is indeed violated. As effective field theory which includes 
all features of the standard model as well as all possible Lorentz violating terms 
was proposed by Kostelecky and Mewes and goes by the name Effective Standard-Model 
Extension (SME) \cite{Kostelecky:2003xn,Kostelecky:2003cr}.
However, most studies on neutrino oscillations use  a more phenomenological approach 
wherein a few Lorentz violating terms are added to the standard model Lagrangian. 
For example, many studies add to the standard Lagrangian an effective 
CPT  violating term parametrised as
%\be
%{\cal L} = i \bar \psi \partial_\mu \gamma^\mu\psi - m\bar\psi\psi 
%-A_\mu  \bar \psi  \gamma^\mu\psi  - B_\mu  \bar \psi \gamma_5\gamma^\mu\psi 
%\label{eq:liv}
%\ee
%where the quantities $A_\mu$ and $B_\mu$ are just real numbers and hence 
%bring about the breaking of Lorentz invariance in the neutrino sector.  This breaking of Lorentz invariance then leads to CPT violation for the group of proper Lorentz transformation and 
%the relevant part of the Lagrangian is given by
\be
{\cal L}_\nu^{\rm CPTV} = \bar\nu_L^\alpha b^\mu_{\alpha\beta} \gamma_\mu \nu_L^\beta
\label{eq:livnu}
\ee
where the Lorentz and CPT is explicitly seen to be violated by the $3\times 3$ 
Hermitian matrices which carry the bare Lorentz index $\mu$  and flavor indices 
$\alpha$ and $\beta$.  The only surviving CPTV component is $b^0_{\alpha\beta}$. 
The form of this term is similar to that of the matter potential and results in changing the 
dispersion relation of the neutrinos in vacuum to
\be
H = \frac{MM^\dagger}{2p} + b^0
\ee
where $M$ is the neutrino mass matrix and $b^0$ is a $3\times 3$ non-diagonal matrix. 
Atmospheric neutrinos also travel 
in matter and therefore the neutrino Hamiltonian in matter in presence of CPTV 
written in the flavor basis 
is given by
\be
H_f = \frac{1}{2E} U_{PMNS} diag(0,\ms,\ma)U_{PMNS}^\dagger 
+ U_b diag(0,\delta b_{21},\delta b_{31})U_b^\dagger + diag(A,0,0)
\ee
where $\delta b_{ij} = b_i - b_j$ and 
$U_b$ is the matrix which diagonalises $b^0$ 
giving eigenvalues $b_i$, and the other quantities are as defined in the 
previous sections. The mixing matrix $U_b$ has 3 angles and 3 phases, all of 
which are physical while $U_{PMNS}$ has 3 angles and 1 phase. 
Considering the most general case with these 6 mixing angles, 4 phases and 
4 independent eigenvalues parametrised as $\ms$, $\ma$, $\delta b_{21}$ and 
$\delta b_{31}$ can be challenging. However attempts have been made in the literature 
to study the impact of CPTV on atmospheric neutrinos (see \cite{Chatterjee:2014oda} 
and references therein).

In \cite{Chatterjee:2014oda}, the authors 
studied the effect of CPTV in atmospheric signals in detectors like INO which have 
charge identification capabilities. The possibility of separating the neutrino from anti-neutrino 
signals at these detectors give them an added handle in containing these CPTV 
parameters. 
%Fig. \ref{fig:cptv1} shows the impact of the CPTV term on 
%muon neutrino survival probability through the difference $\Delta P = P_{\mu\mu}^{U_b\neq
%0} - P_{\mu\mu}^{U_b = 0}$ for $\delta b_{31}=3\times 10^{-23}$ GeV and 2 values of 
%$\theta_{b13}$. 
From a $\chi^2$ analysis of the expected 10 years data from INO, 
the authors \cite{Chatterjee:2014oda} determine that 
INO could restrict $\delta b_{31} \gtap 4\times 10^{-23}$ 
GeV at the 99\% C.L. for both types of neutrino mass hierarchies, while the constraints 
on $\delta b_{21}$ are not competitive with other experiments.

\section{conclusions}

Atmospheric neutrinos detected in SK 
were the first to give unambiguous evidence of neutrino 
oscillations and hence the first solid evidence for physics beyond the standard model. 
This landmark achievement was acknowledged by honouring 
Prof. Kajita of the SK experiment with the Nobel Prize in 2015. The data collected at 
SK still is a driving force in global analyses. It also provides some hints regarding the 
as-of-yet unknowns in neutrino physics, such as the octant of $\theta_{23}$, 
$\delta_{CP}$ and to a small extent the mass hierarchy. All of these parameters 
play a sub-dominant role in the three-generation oscillation probabilities relevant for 
atmospheric neutrinos. 

A variety of next generation atmospheric neutrinos detectors such as the HK, 
PINGU, ORCA and INO have to been proposed to catch some of these sub-leading 
aspects, particularly the neutrino mass hierarchy. In this article, we have discussed the 
expected reach of these future experiments towards discovery of the true mass hierarchy and 
octant of $\theta_{23}$. We have also discussed the importance of atmospheric neutrinos in 
probing new physics such as presence of sterile neutrinos, NSI and CPT violation. 
Atmospheric neutrino experiments indeed could continue to play a crucial role in the 
field of neutrino physics in the years to come.

\begin{acknowledgments}
SC acknowledge
support from the Neutrino Project under the XII plan of Harish-Chandra
Research Institute and partial support from the European 
Union FP7 ITN INVISIBLES (Marie Curie Actions, PITN-GA-2011-289442)
\end{acknowledgments}

\bibliography{references}

\end{document}